# Anisotropic mechanical properties and strain tuneable band-gap in single-layer SiP, SiAs, GeP and GeAs

Bohayra Mortazavi[*,1] and Timon Rabczuk[#,2]

[1]*Institute of Structural Mechanics, Bauhaus-Universität Weimar, Marienstr. 15, D-99423 Weimar, Germany.*

[2]*College of Civil Engineering, Department of Geotechnical Engineering, Tongji University, Shanghai, China.*

**Abstract**

Group IV–V-type two-dimensional (2D) materials, such as GeP, GeAs, SiP and SiAs with anisotropic atomic structures, have recently attracted remarkable attention due to their outstanding physics. In this investigation, we conducted density functional theory simulations to explore the mechanical responses of these novel 2D systems. In particular, we explored the possibility of band-gap engineering in these 2D structures through different mechanical loading conditions. First-principles results of uniaxial tensile simulations confirm anisotropic mechanical responses of these novel 2D structures, with considerably higher elastic modulus, tensile strength and stretchability along the zigzag direction as compared with the armchair direction. Notably, the stretchability of considered monolayers along the zigzag direction was found to be slightly higher than that of the single-layer graphene and h-BN. The electronic band-gaps of energy minimized single-layer SiP, SiAs, GeP and GeAs were estimated by HSE06 method to be 2.58 eV, 2.3 eV, 2.24 eV and 1.98 eV, respectively. Our results highlight the strain tuneable band-gap character in single-layer SiP, SiAs, GeP and GeAs and suggest that various mechanical loading conditions can be employed to finely narrow the electronic band-gaps in these structures.

*Corresponding author (Bohayra Mortazavi):  bohayra.mortazavi@gmail.com
Tel: +49 157 8037 8770; Fax: +49 364 358 4511; [#]Timon.rabczuk@uni-weimar.de

1. Introduction

During the last decade, the interest in the two-dimensional (2D) materials has kept continuously increasing, by theoretical predictions and experimental realization of new members in this family. 2D materials family comprise both isotropic members with high lattice symmetry, such as graphene [1,2], and some low-symmetry anisotropic members, such as 1T' transition metal dichalcogenides [3–5] and distorted-1T rhenium disulfide [6]. Although the 2D materials with high symmetry lattices have been to-date considerably more successful in attracting the interest of scientific community, in recent years the low-symmetry 2D lattices are gaining remarkable attentions [7]. The recent interests toward the anisotropic 2D materials originate from their unique optical (linear and nonlinear), electrical, mechanical, and thermoelectric properties; which create outstanding possibilities for the design of novel angle-dependent devices, including polarization-sensitive photodetectors, integrated digital inverters, mid-infrared polarizers, artificial synaptic devices, linearly polarized ultrafast lasers, and polarization sensors [7–10]. It is quiet well-known that the fundamental properties of 2D materials strongly correlate to the crystal structure and symmetry. This issue highlights the importance of exploring different aspects of in-plane anisotropy in 2D materials atomic lattices in order to provide understanding of the anisotropic structure–property relations.

Group IV–V-type 2D materials, such as germanium phosphide (GeP), germanium arsenide (GeAs), silicon phosphide (SiP) and silicon arsenide (SiAs) are another family of low-symmetry materials, predicted originally by Ashton *et al.* [11] in 2016. The bulk structures of GeP, GeAs, SiP and SiAs are quite well-known to exhibit anisotropic material properties. Bulk GeP and GeAs structures have been experimentally realized by Donohue and Yang [12], back to 1970. SiP and SiAs bulk layered structures were also studied by Beck and Stickler [13]. Recently, bulk single crystals of GeP, GeAs, SiP and SiAs have been successfully grown by melt-growth under high pressure in a cubic anvil hot press [14]. As an exciting matter of fact concerning these bulk structures, due to the weak van der Waals interlayer interactions, experimental realization of their 2D structures can be accomplished efficiently by exfoliation. In this regard, most recently GeAs nanomembranes were experimentally realized by Yang *et al.* [7], which were found to yield highly anisotropic in-plane electronic and optical properties. In another most recent experimental advance, for the first time 2D GeP structures were fabricated by Li *et* *al.* [15], with strong in-plane anisotropic physical properties. Worthy to note that the thermal and dynamical stability of single-layer GeP and GeAs have been recently studied theoretically by Zhou *et al.* [16]. Recent exciting experimental advances with respect to the synthesis of 2D GeP and GeAs, certainly highlight the practical prospect for the experimental realization of SiP and SiAs in 2D form in the near future. Worth mentioning that in a recent first-principles density functional theory (DFT) investigation by Cheng *et al.* [17], the exfoliation energy for SiP, SiAs, GeP and GeAs were predicted to be 0.26 J/m$^2$, ~0.27 J/m$^2$, 0.34 J/m$^2$ and 0.37 J/m$^2$, respectively. Interestingly, the exfoliation energy of SiP and SiAs are lower than that of the graphite (0.32 J/m$^2$), which further highlight the experimental prospects for their monolayer realization. To the best of our knowledge, mechanical properties of these novel 2D structures have been studied neither theoretically nor experimentally. The objective of present study is therefore to explore the mechanical properties as well as the strain engineering of electronic properties of single-layer GeP, GeAs, SiP and SiAs through using the first-principles DFT simulations.

## 2. Computational method

In order to explore the mechanical and electronic properties of single-layer GeP, GeAs, SiP and SiAs, first-principles DFT simulations were conducted using the Vienna ab-initio simulation package (VASP) [18–20]. The plane wave basis set with an energy cut-off of 500 eV and generalized gradient approximation (GGA) exchange-correlation functional proposed by Perdew-Burke-Ernzerhof (PBE) [21] were also employed. Periodic boundary conditions were applied in all directions with a vacuum layer of 20Å to avoid the image-image interactions along the sheet thickness. VESTA [22] package was employed in order to illustrate the structures. To evaluate the mechanical properties, we conducted uniaxial tensile modelling for a unit-cell. To this aim, we increased the size of the periodic simulation box along the loading direction with a constant engineering strain step. We remind that for the uniaxial loading of 2D materials, upon the stretching along the loading direction, the stress along the transverse direction should stay negligible. To satisfy this condition, after applying the loading strain, the simulation box size along the transverse direction of the loading was changed accordingly in a way that the transverse stress remains negligible. After applying the changes in the simulation box size, the atomic positions were rescaled to avoid any sudden void formation or bond stretching as well. We then used the conjugate gradient method for the geometry optimizations, with strict termination criteria of $10^{-5}$ eV and 0.01 eV/Å for the energy and forces, respectively, within the tetrahedron method with Blöchl corrections [23]. Since the atomic lattices of studied monolayers are highly elongated in one direction (armchair direction as shown in Fig, 1), we used 3×7×1 Monkhorst-Pack [24] k-point mesh size, in which the lower k-point mesh of 3 was used along the direction with the longer length (armchair direction). Since the PBE functional are well-known to underestimate the band-gap values, we also employed the screened hybrid functional, HSE06 [25] to investigate the electronic band-gap, with the same K-point mesh size as that was used for the evaluation of mechanical properties.

## 3. Results and discussions

We first employed the DFT calculations for the unit-cells energy minimization and corresponding geometry optimization. Fig.1, illustrate the top and side views of atomic lattices of energy minimized single-layer SiP, SiAs, GeP and GeAs, which show ABC atomic stacking sequence. In order to analyse the anisotropicity in mechanical properties, we evaluated the mechanical response along the armchair and zigzag directions, as depicted in Fig. 1. The lattice constants of energy minimized monolayers along the armchair and zigzag are summarized in Table 1.

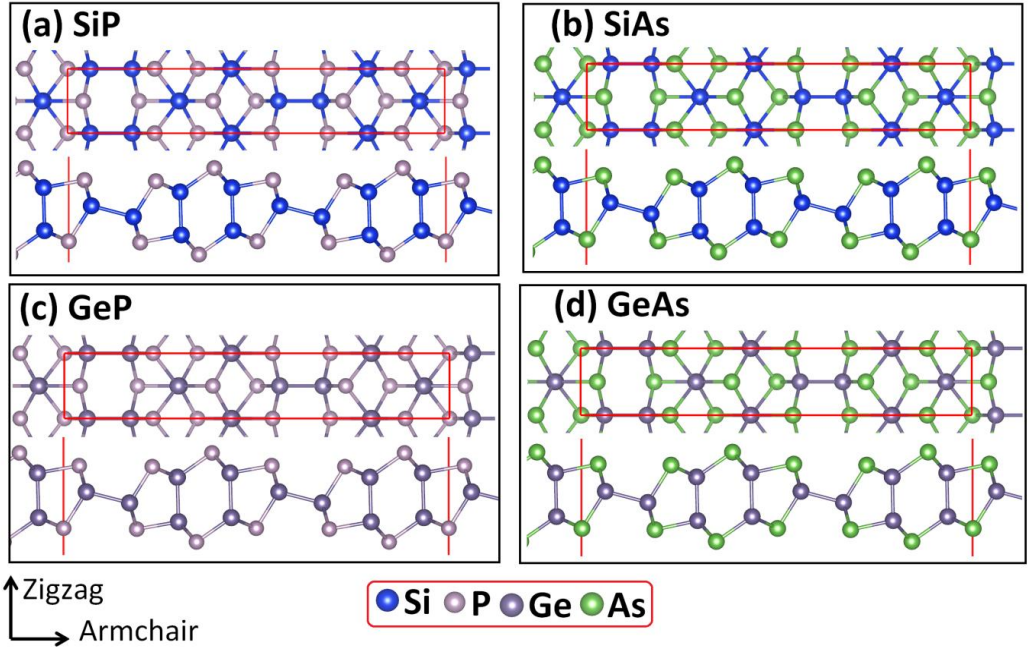

**Fig. 1**, Top and side views of atomic configuration in single-layer SiP, SiAs, GeP and GeAs.

Table. 1, Lattice constants of energy minimized single-layer SiP, SiAs, GeP and GeAs.

| Structure | Lattice length (Å) | |
| --- | --- | --- |
| | armchair | zigzag |
| SiP | 20.538 | 3.532 |
| SiAs | 21.324 | 3.696 |
| GeP | 21.492 | 3.662 |
| GeAs | 22.252 | 3.821 |

In Fig. 2, the DFT predictions for the uniaxial stress-strain responses of single-layer SiP, SiAs, GeP and GeAs elongated along the armchair and zigzag directions are compared. In all cases, the stress-strain responses present an initial linear relation which is followed by a nonlinear trend up to the ultimate tensile strength point. The slope of the first initial linear section of stress-strain curve is equal to the elastic modulus. In this work, we therefore fitted lines to the stress-strain values for the strain levels below ~0.02 to report the elastic modulus. Within the elastic region, the strain along the traverse direction ($\varepsilon_t$) with respect to the loading strain ($\varepsilon_l$) is constant and can be used to evaluate the Poisson's ratio, as: $-\varepsilon_t/\varepsilon_l$. Our results for the uniaxial loading along the armchair and zigzag directions shown in Fig. 2, for the all considered 2D structures clearly reveal that the both linear and nonlinear sections of stress-strain responses are different. In Table 2, the mechanical properties of these novel 2D structures are summarized.

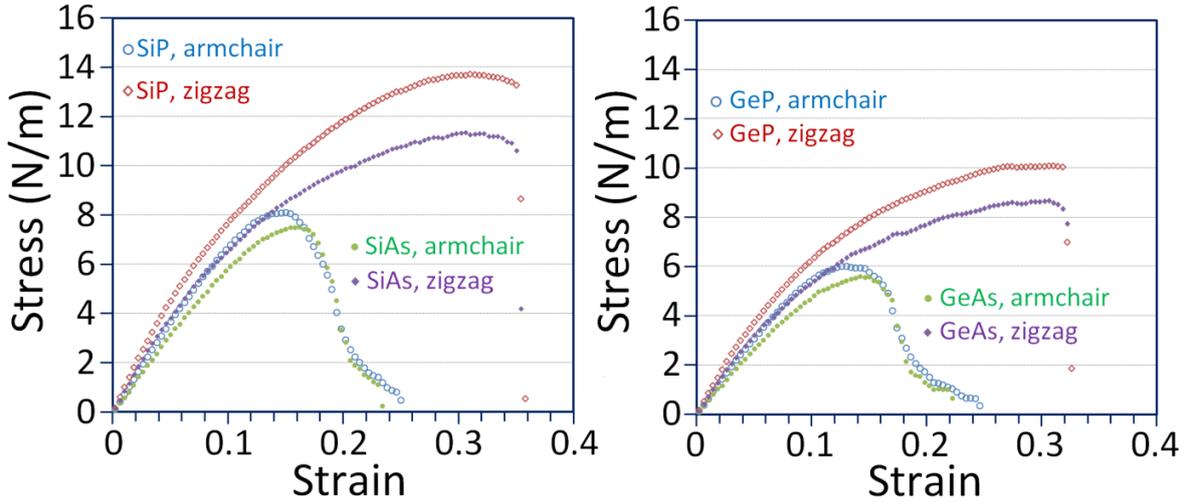

Fig. 2, Uniaxial stress-strain responses of single-layer and free-standing SiP, SiAs, GeP and GeAs stretched along the armchair and zigzag directions.

Interestingly, along the zigzag direction these systems show considerably higher elastic modulus, tensile strength and strain at ultimate tensile strength as well. These results confirm the highly anisotropic mechanical response of these 2D materials. As it is clear, single-layer SiP and GeAs, respectively, exhibit the highest and lowest, elastic modulus and tensile strengths. Interestingly, the strain at ultimate tensile strength which is the representative of the stretchability of these structures were found to be close and convincingly independent of the atomic compositions. Notably, along the zigzag direction the stretchability of studied monolayers is by around two times higher than that along the armchair direction. As an interesting finding, the stretchability of single-layer SiP, SiAs, GeP and GeAs along the zigzag are slightly higher than that of the high-symmetry 2D structures. Worthy to remind that the strain at the ultimate tensile strength point for pristine graphene and hexagonal boron-nitride were found to be ~0.27 [26] and ~0.3 [27], respectively, which are lower than the value of ~0.31 that we predicted for the uniaxial loading of these novel 2D structures along the zigzag.

Table 2, Summarized mechanical properties of single-layer SiP, SiAs, GeP and GeAs along the armchair and zigzag directions. $Y$, $P$, $UTS$ and $SUTS$ stand for elastic modulus, Poisson's ratio, ultimate tensile strength and strain at ultimate tensile strength point, respectively. Stress units are in N/m.

| Structure | $Y_{armchair}$ | $Y_{zigzag}$ | $P_{armchair}$ | $P_{zigzag}$ | $UTS_{armchair}$ | $UTS_{zigzag}$ | $SUTS_{armchair}$ | $SUTS_{zigzag}$ |
|---|---|---|---|---|---|---|---|---|
| SiP | 77.2 | 100.4 | 0.14 | 0.14 | 8.1 | 13.7 | 0.15 | 0.31 |
| SiAs | 64.7 | 86.3 | 0.11 | 0.16 | 7.5 | 11.4 | 0.16 | 0.31 |
| GeP | 65.6 | 82.1 | 0.15 | 0.16 | 6.0 | 10.1 | 0.15 | 0.31 |
| GeAs | 60 | 72.2 | 0.16 | 0.16 | 5.6 | 8.7 | 0.15 | 0.31 |

In order to better understand the bonding nature in these novel 2D systems, in Fig. 3 the electron localization function (ELF) [28] 3D profiles are plotted. The ELF is a position-dependent function ranging from 0 to 1. The ELF values close to one corresponds to the region with high probability of finding electron localizations and ELF = 0.5 corresponds to the region of electron gas-like behaviour. For the studied monolayers, the ELF values around the center of all bonds are greater than 0.75, confirming the covalent bonding in these structures. Nevertheless, the electron localization is also considerable around the As and P atoms originated from their higher valance electrons as well as their higher electronegativity leading to charge gain from Ge and Si atoms. This charge transfer also induce ionic interactions between heteronuclear bonds in these 2D systems.

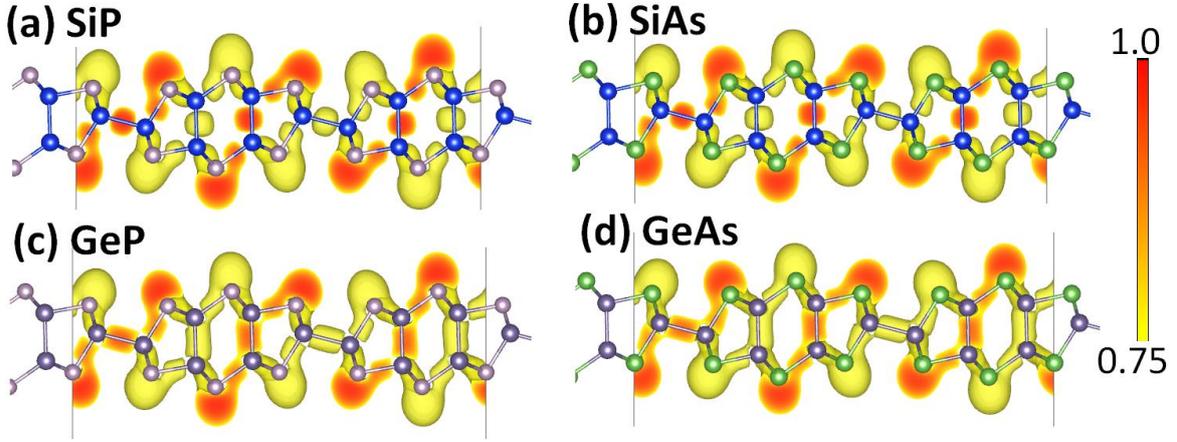

**Fig. 3**, The side views of single-layer SiP, SiAs, GeP and GeAs along with their electron localization function 3D profiles.

According to our analysis of deformation process, the lower tensile strength and stretchability along the armchair direction for these 2D structures can be explained due to the Si-Si and Ge-Ge homonuclear bonds. These homonuclear bonds are exactly oriented along the armchair direction, and for the loading along this direction they involve directly in the load bearing and stretching as well. These homonuclear bonds are also weaker in comparison with heteronuclear bonds, as they are fully covalent and lack the ionic contributions. As observable in Fig. 1, around these homonuclear bonds the packing density of studied 2D materials are also minimum. It was found that the for the uniaxial loading along the armchair direction these bond exhibit the highest elongation and the final structural rupture also occurs in these bonds. For the uniaxial loading along the zigzag direction, these homonuclear bonds are exactly oriented along the transverse direction of loading and thus their softer stiffness cannot suppress the mechanical properties. These findings reveal the interesting deformation mechanism of studied monolayers which are different from other 2D materials [29–38].

We next probe the electronic properties of single-layer SiP, SiAs, GeP and GeAs by calculating the total electronic density of states (DOS) within the PBE and HSE06 methods. In Fig. 4, the acquired DOSs for the stress-free single-layer SiP, SiAs, GeP and GeAs predicted by PBE and HSE06 methods are compared. As expected, the PBE results underestimate the band-gap values predicted by the HSE06 method.

According to the PBE method, the band-gaps of free-standing and single-layer SiP, SiAs, GeP and GeAs were found to be 1.82 eV, 1.65 eV, 1.58 eV and 1.32 eV, respectively. The HSE06 method predicts higher band-gap values of 2.58 eV, 2.3 eV, 2.24 eV and 1.98 eV for energy minimized single-layer SiP, SiAs, GeP and GeAs, respectively. The predicted band-gap values are in close with those reported in the recent study by Cheng *et al.* [17].

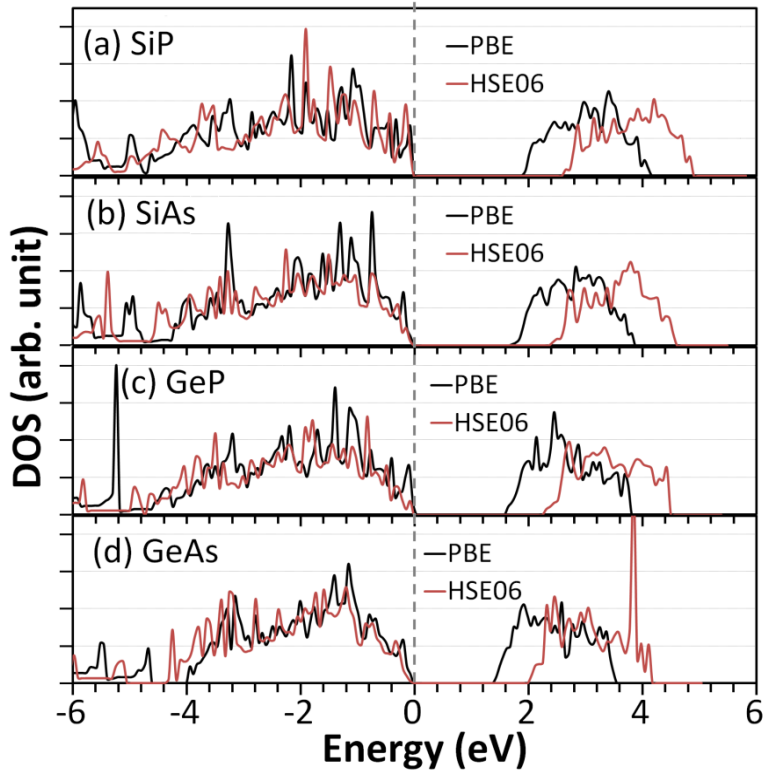

Fig. 4, PBE and HSE06 results for the electronic density of states (DOS) of unstrained single-layer and free-standing SiP, SiAs, GeP and GeAs. The Fermi energy is aligned to zero.

Taking into consideration that the HSE06 method provides more accurate predictions for the band-gap values, in Fig. 5 we used this method to specifically analyse the possibility of band-gap engineering in single-layer SiP, SiAs, GeP and GeAs by various mechanical loading conditions. In this case, we also conducted the biaxial tensile loading simulations and compared the results with those of uniaxial tensile loadings. As shown in Fig. 5a, for the single-layer SiP by uniaxial loading along the zigzag and biaxial loading, the band-gap values were found to be close. In these cases the electronic band-gap almost linearly decreases by increasing the strain value, reaching to ~1.6 eV for the strain level of 0.1. In this structure, for the uniaxial loading along the armchair the band-gap first slightly increases and then decreases, exhibiting a band-gap at strain level of 0.04, very close to that of the unstrained original structure. After this strain level the band-gap keeps decreasing also in an approximately linear pattern with respect to the strain level. As illustrated in Fig. 5b, the estimated band-gap values of SiAs monolayer, from the strain level of 0.04 for the uniaxial loading along the zigzag and biaxial loading are very close and they both decrease linearly, presenting a band-gap of ~1.35 eV for the strain value of 0.1. Likely to the SiP monolayer, for the uniaxial loading of SiAs along the armchair the band-gap first slight increases and then almost linearly decrease and finally reaches a value of ~2 eV for strain value of 0.1.

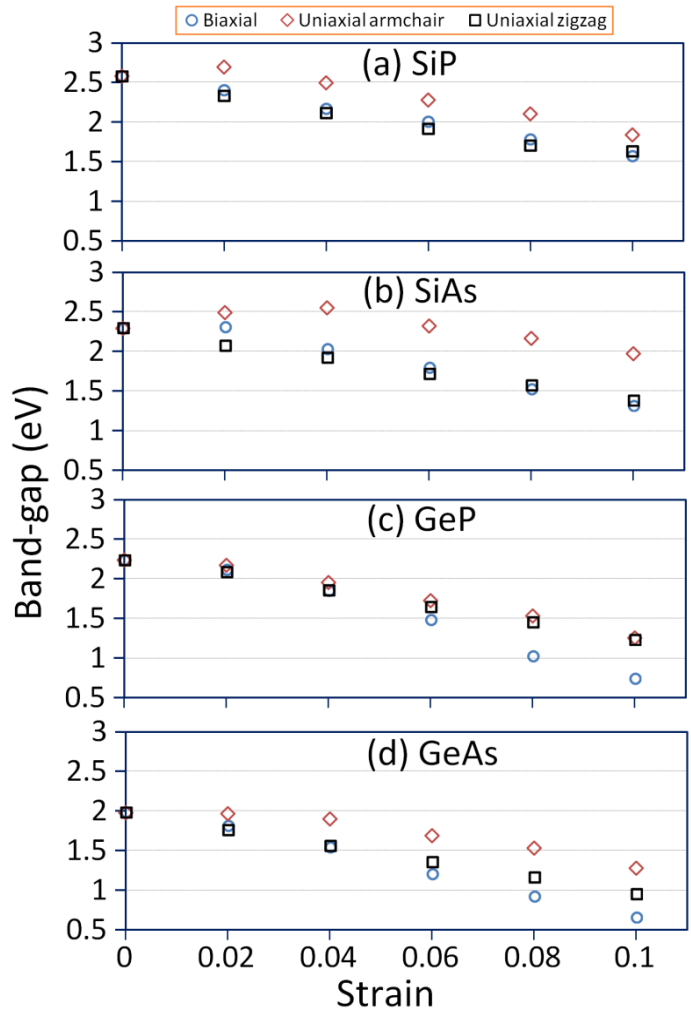

**Fig. 5**, Electronic band-gaps of single-layer SiP, SiAs, GeP and GeAs as a function of applied strain, predicted by the HSE06 functional. The Fermi energy is aligned to zero.

For the single-layer GeP, by applying the mechanical tensile strains the band-gap values decrease for all cases, as clearly observable in Fig. 5c. In this monolayer, by uniaxial loading along the armchair and zigzag the band-gap values change very closely and drop to a value of ~1.25 eV at the strain value of 0.1. For this 2D structure, the band-gap value can be more considerably tuned by the biaxial loading, dropping to ~0.75 eV for the biaxial strain of 0.1. Among the all studied monolayers, GeAs was found to yield the minimum band-gap values upon the mechanical straining. In this structure as depicted in Fig. 5d, the biaxial loading and uniaxial loading along the armchair, respectively, yield the maximum and minimum effects on the band-gap engineering. For the biaxial strain of 0.1, the band-gap of single-layer GeAs was predicted to drop to 0.66 eV, whereas for the same strain level and for the uniaxial loading along the armchair, the electronic band-gap was estimated to be almost twice, 1.28 eV. Our results highlight the strain tuneable band-gap character in single-layer and free-standing SiP, SiAs, GeP and GeAs and confirm that various mechanical loading conditions can be employed to finely alter the electronic responses of these novel 2D structures. The obtained results however suggest the limited chance for the further band-gap opening in these monolayers upon the mechanical straining. Worthy to note that the narrowing of the band-gap in single-layer SiP, SiAs, GeP and GeAs upon the mechanical loading is a very attractive finding in order to substitute the silicon with band-gap of 1.1 eV in electronic devices. We would like to remind that that in the most cases, the 2D materials are in multi-layer forms and that may affect their electronic properties. Therefore the analysis of effects of thickness on the electronic and optical properties of studied 2D materials can be an important and attractive topic for the future studies.

## 4. Conclusion

Group IV–V-type 2D materials, such as germanium phosphide (GeP), germanium arsenide (GeAs), silicon phosphide (SiP) and silicon arsenide (SiAs) are a novel family of low-symmetry 2D structures, which have recently garnered growing attention stemming from their outstanding optical and electrical properties. Motivated by the recent experimental advances in the fabrication of GeP and GeAs in 2D forms, we conducted first-principles density functional theory simulations to explore the mechanical responses and electronic band-gap of single-layer SiP, SiAs, GeP and GeAs. We first studied the mechanical properties by performing the uniaxial tensile simulations. Our first-principles results confirm the highly anisotropic mechanical responses along the all considered monolayers, in which along the zigzag direction these systems show considerably higher elastic modulus, tensile strength and stretchability as well. As an interesting finding, the stretchability of single-layer

SiP, SiAs, GeP and GeAs along the zigzag were found to be slightly higher than that of the high-symmetry 2D structures, such as the graphene and hexagonal boron-nitride. The lower tensile strength as well as the stretchability along the armchair direction was found to be mainly due to the softening effects of Si-Si or Ge-Ge homonuclear bonds, which are exactly oriented along the armchair direction. According to the HSE06 method results, the electronic band-gaps of energy minimized single-layer SiP, SiAs, GeP and GeAs were estimated to be 2.58 eV, 2.3 eV, 2.24 eV and 1.98 eV, respectively. Our results highlight the strain tuneable band-gap character in single-layer SiP, SiAs, GeP and GeAs and confirm that various mechanical loading conditions can be employed to finely alter the electronic responses of these novel 2D materials.

## Acknowledgment

Authors greatly acknowledge the financial support by European Research Council for COMBAT project (Grant number 615132).

# Supporting information

# Anisotropic mechanical properties and strain tuneable band-gap in single-layer SiP, SiAs, GeP and GeAs

Bohayra Mortazavi[*,1] and Timon Rabczuk[2]

[1]*Institute of Structural Mechanics, Bauhaus-Universität Weimar, Marienstr. 15, D-99423 Weimar, Germany.*

[2]*College of Civil Engineering, Department of Geotechnical Engineering, Tongji University, Shanghai, China.*

*Corresponding author (Bohayra Mortazavi):  bohayra.mortazavi@gmail.com

## Atomic lattices of single-layer SiP, SiAs, GeP and GeAs in VASP POSCAR

### (a) SiP

```
SiP
   1.00000000000000
     3.5323553796044052    0.0000000000000000    0.0000000000000000
     0.0000000000000000   20.5376509368007909    0.0000000000000000
     0.0000000000000000    0.0000000000000000   20.0000000000000000
   Si   P
   12   12
Direct
  0.0000000000000000  0.1711094682363097  0.5770656506239220
  0.5000002980000033  0.6711094682363097  0.5770656506239220
  0.0000000000000000  0.0598695890787511  0.6071275181045905
  0.5000001780000005  0.5598696000787555  0.6071275181045905
  0.9999996419999988  0.4356792884266198  0.5391805126690983
  0.4999999099999997  0.9356792594266068  0.5391805126690983
  0.0000000000000000  0.4327858341840667  0.6568007291607856
  0.5000003580000012  0.9327858641840621  0.6568007291607856
  0.5000005959999996  0.2953134699145323  0.6450875816419739
  0.0000007860000011  0.7953134399145370  0.6450875816419739
  0.5000001200000028  0.2983177088409334  0.5274306305188077
  0.0000003099999972  0.7983177088409334  0.5274306305188077
  0.5000007160000024  0.4999503068710283  0.5098379326059188
  0.0000008660000006  0.9999503368710307  0.5098379326059188
  0.9999995820000009  0.2310124280211738  0.6744469455864959
  0.4999997919999970  0.7310124430211786  0.6744469455864959
  0.0000000000000000  0.3451117040676976  0.4736877183230135
  0.5000003580000012  0.8451117040676976  0.4736877183230135
  0.4999994940000008  0.0416192409123823  0.6775626076018000
  0.9999996419999988  0.5416192669123774  0.6775626076018000
  0.5000000599999979  0.1895337471902394  0.5066559696305148
  0.0000002680000009  0.6895337471902323  0.5066559696305148
  0.5000001200000028  0.3859736752562952  0.7104867469025606
  0.0000003440000000  0.8859736452562927  0.7104867469025606
```

## (b) SiAs

```
SiAs
   1.00000000000000
     3.6960200901533051    0.0000000000000000    0.0000000000000000
     0.0000000000000000   21.3238268224045413    0.0000000000000000
     0.0000000000000000    0.0000000000000000   20.0000000000000000
   Si   As
    12   12
Direct
  0.0000000000000000   0.1687332190751007   0.5765308992914555
  0.5000002980000033   0.6687332190751007   0.5765308992914555
  0.0000000000000000   0.0623089399147361   0.6075915428639433
  0.5000001780000005   0.5623089509147476   0.6075915428639433
  0.9999996419999988   0.4369162307316756   0.5389616292351036
  0.4999999099999997   0.9369162017316697   0.5389616292351036
  0.0000000000000000   0.4347827672375928   0.6563404611060335
  0.5000003580000012   0.9347827972376024   0.6563404611060335
  0.5000005959999996   0.2939786298427336   0.6454034209967503
  0.0000007860000011   0.7939785998427453   0.6454034209967503
  0.5000001200000028   0.2963451067870508   0.5280400747359835
  0.0000003099999972   0.7963451067870650   0.5280400747359835
  0.5000007160000024   0.5019300712932093   0.5049234113218830
  0.0000008660000006   0.0019301012932118   0.5049234113218830
  0.9999995820000009   0.2290029335311345   0.6794013743509950
  0.4999997919999970   0.7290029485311393   0.6794013743509950
  0.0000000000000000   0.3443685144608253   0.4703991314722131
  0.5000003580000012   0.8443685144608111   0.4703991314722131
  0.4999994940000008   0.0444819751305019   0.6830240470400071
  0.9999996419999988   0.5444820011304969   0.6830240470400071
  0.5000000599999979   0.1868192586604067   0.5010984059140071
  0.0000002680000009   0.6868192586604067   0.5010984059140071
  0.5000001200000028   0.3866088143349486   0.7136561450410710
  0.0000003440000000   0.8866087843349391   0.7136561450410710
```

## (c) GeP

```
GeP
   1.00000000000000
     3.6621363765471240    0.0000000000000000    0.0000000000000000
     0.0000000000000000   21.4917188838344586    0.0000000000000000
     0.0000000000000000    0.0000000000000000   20.0000000000000000
   Ge   P
    12   12
Direct
  0.0000000000000000   0.1716289769114709   0.5766317713483744
  0.5000002980000033   0.6716289769114780   0.5766317713483744
  0.0000000000000000   0.0593349213559691   0.6075983198032233
  0.5000001780000005   0.5593349323559735   0.6075983198032233
  0.9999996419999988   0.4356028214164880   0.5355290302624169
  0.4999999099999997   0.9356027924164820   0.5355290302624169
  0.0000000000000000   0.4324468375774089   0.6605447355717970
  0.5000003580000012   0.9324468675774042   0.6605447355717970
  0.5000005959999996   0.2955271531939871   0.6486655339978498
  0.0000007860000011   0.7955271231939918   0.6486655339978498
  0.5000001200000028   0.2986004015676968   0.5237338402643417
  0.0000003099999972   0.7986004015677040   0.5237338402643417
  0.5000007160000024   0.5004238183994900   0.5056365040077750
  0.0000008660000006   0.0004238483994925   0.5056365040077750
  0.9999995820000009   0.2306844476226075   0.6785039381714455
  0.4999997919999970   0.7306844626226052   0.6785039381714455
```

```
  0.0000000000000000  0.3453962379638895  0.4671745295547041
  0.5000003580000012  0.8453962379638895  0.4671745295547041
  0.4999994940000008  0.0406944561654825  0.6812664115546951
  0.9999996419999988  0.5406944821654776  0.6812664115546951
  0.5000000599999979  0.1903106508092236  0.5030645834539484
  0.0000002680000009  0.6903106508092236  0.5030645834539484
  0.5000001200000028  0.3856257380163015  0.7170213453789671
  0.0000003440000000  0.8856257080163061  0.7170213453789671
```

### (d) GeAs

```
GeAs
   1.00000000000000
     3.8214126082976501    0.0000000000000000    0.0000000000000000
     0.0000000000000000   22.2520843572523717    0.0000000000000000
     0.0000000000000000    0.0000000000000000   20.0000000000000000
   Ge   As
   12   12
Direct
  0.0000000000000000  0.1698600876681766  0.5760089632278209
  0.5000002980000033  0.6698600876681766  0.5760089632278209
  0.0000000000000000  0.0613383790450470  0.6078570844353735
  0.5000001780000005  0.5613383900450444  0.6078570844353735
  0.9999996419999988  0.4363819217938456  0.5356861161366453
  0.4999999099999997  0.9363818927938468  0.5356861161366453
  0.0000000000000000  0.4340301654476306  0.6603685872617859
  0.5000003580000012  0.9340301954476260  0.6603685872617859
  0.5000005959999996  0.2945384471763859  0.6488707254463364
  0.0000007860000011  0.7945384171763905  0.6488707254463364
  0.5000001200000028  0.2969917605952901  0.5239960548696203
  0.0000003099999972  0.7969917605952901  0.5239960548696203
  0.5000007160000024  0.5015330608410693  0.5015569092691337
  0.0000008660000006  0.0015330908410647  0.5015569092691337
  0.9999995820000009  0.2292707708871475  0.6828741972632031
  0.4999997919999970  0.7292707858871523  0.6828741972632031
  0.0000000000000000  0.3449171825113595  0.4644021985694238
  0.5000003580000012  0.8449171825113595  0.4644021985694238
  0.4999994940000008  0.0429241120820762  0.6857704304649559
  0.9999996419999988  0.5429241380820713  0.6857704304649559
  0.5000000599999979  0.1881855681981364  0.4979994173010738
  0.0000002680000009  0.6881855681981364  0.4979994173010738
  0.5000001200000028  0.3863050047538366  0.7199798591241020
  0.0000003440000000  0.8863049747538341  0.7199798591241020
```